\documentclass[aps,prb,twocolumn,showpacs,groupedaddress]{revtex4-1}
\usepackage{graphics}
\usepackage{graphicx}
\pdfoutput=1
\usepackage{dcolumn}
\usepackage{bm}
\usepackage{float}
\usepackage{stmaryrd}
\usepackage{wasysym}
\usepackage{pifont}

\begin{document}
\title{Thermoelectric transport properties of CaMg$_2$Bi$_2$, EuMg$_2$Bi$_2$, and YbMg$_2$Bi$_2$}

\author{Andrew F. May}
\email{mayaf@ornl.gov}
\author{Michael A. McGuire}
\affiliation{Materials Science and Technology Division, Oak Ridge National Laboratory, Oak Ridge, TN 37831}
\author{Jie Ma}
\affiliation{Quantum Condensed Matter Division, Oak Ridge National Laboratory, Oak Ridge, TN 37831}
\author{Olivier Delaire}
\affiliation{Quantum Condensed Matter Division, Oak Ridge National Laboratory, Oak Ridge, TN 37831}
\author{David J. Singh}
\affiliation{Materials Science and Technology Division, Oak Ridge National Laboratory, Oak Ridge, TN 37831, USA}
\author{Wei Cai}
\affiliation{High Temperature Materials Laboratory, Oak Ridge National Laboratory, Oak Ridge, TN 37831}
\author{Ashfia Huq}
\affiliation{Quantum Condensed Matter Division, Oak Ridge National Laboratory, Oak Ridge, TN 37831}
\author{Hsin Wang}
\affiliation{High Temperature Materials Laboratory, Oak Ridge National Laboratory, Oak Ridge, TN 37831}
\date{\today}

\begin{abstract}
The thermoelectric transport properties of CaMg$_2$Bi$_2$, EuMg$_2$Bi$_2$, and YbMg$_2$Bi$_2$ were characterized between 2 and 650\,K.  As synthesized, the polycrystalline samples are found to have lower $p$-type carrier concentrations than single-crystalline samples of the same empirical formula.  These low carrier concentration samples possess the highest mobilities yet reported for materials with the CaAl$_2$Si$_2$ structure type, with a mobility of $\sim$740\,cm$^2$/V/s observed in EuMg$_2$Bi$_2$ at 50\,K.  Despite decreases in the Seebeck coefficient ($\alpha$) and electrical resistivity ($\rho$) with increasing temperature, the power factor ($\alpha^2/\rho$) increases for all temperatures examined.  This behavior suggests a strong asymmetry in the conduction of electrons and holes.  The highest figure of merit ($zT$) is observed in YbMg$_2$Bi$_2$, with $zT$ approaching 0.4 at 600\,K for two samples with carrier densities of approximately 2$\times$10$^{18}$cm$^{-3}$ and 8$\times$10$^{18}$cm$^{-3}$ at room temperature.  Refinements of neutron powder diffraction data yield similar behavior for the structures of CaMg$_2$Bi$_2$ and YbMg$_2$Bi$_2$, with smooth lattice expansion and relative expansion in $c$ being $\sim$35\% larger than relative expansion in $a$ at 973\,K.  First principles calculations reveal an increasing band gap as Bi is replaced by Sb then As, and subsequent Boltzmann transport calculations predict an increase in $\alpha$ for a given $n$ associated with an increased effective mass as the gap opens.  The magnitude and temperature dependence of $\alpha$ suggests higher $zT$ is likely to be achieved at larger carrier concentrations, roughly an order of magnitude higher than those in the current polycrystalline samples, which is also expected from the detailed calculations.
\end{abstract}

\maketitle

\section{Introduction}

Ternary antimonides with the CaAl$_2$Si$_2$ structure-type ($AX_2$Sb$_2$; $X$=Cd,Zn; $A$=Ca,Yb,Eu) have shown promising thermoelectric performance in the 500 to 800\,K temperature range.  Thermoelectric efficiency is generally discussed in terms of the materials--level figure of merit, $zT=(\alpha^2T)/(\rho\kappa)$, where $\alpha$ is the Seebeck coefficient, $\rho$ the electrical resistivity, and $\kappa$ the thermal conductivity.  The highest figure of merit reported within this family of materials is $zT$ = 1.2 at 700\,K in YbCd$_{1.6}$Zn$_{0.4}$Sb$_2$,\cite{Cd122Grin09} while $zT$ values between 0.5 and 1.1 are typical.\cite{122Jeff,EuZn2Sb2,Cd122Grin10,Cd122Grin10b,YbCdMnSb2,EuCdZnSb2,122Tober}  In general, ternary antimonides that are understood using Zintl chemistry have received significant attention within the thermoelectric community due to their tunability and structural complexity.\cite{KleinkeChemMatReview,ToberAl,Ba4In8Sb16,526Tober,BaZn2Sb2,Ca3AlSb3,MadsenSearch,LiZnSb,BaGa2Sb2,SrZnSb2} However, the analogous bismuthides have received much less attention.

This work is motivated by electrical resistivity and Hall effect measurements on $A$Mg$_2$Bi$_2$ single crystals that revealed properties similar to those in $A$Zn$_2$Sb$_2$.\cite{AMg2Bi2_Inorg}  While nominally charge balanced and predicted to be narrow gap semiconductors,\cite{122Tober,XPS_Yb122_Espen,AMg2Bi2_Inorg} these materials possess $p$-type conduction with carrier densities generally on the order of 10$^{19}$--10$^{20}$cm$^{-3}$, though semiconducting behavior was reported for CaCd$_2$Sb$_2$.\cite{Cd122Grin10}  Interestingly, the rare-earth containing compounds are reported to have higher carrier mobility than the alkaline-earth containing compounds,\cite{AMg2Bi2_Inorg,122Tober} leading to larger $zT$ due to a reduced $\rho$ for a given $\alpha$.  This trend in mobility was observed in single crystalline samples of the compounds studied here. Transport measurements in the \textit{ab}--plane revealed the single crystals to be heavily defective, and the residual resistivity ratios $\rho(300$\,K)/$\rho(10$\,K) were less than 1.6 for all materials, with the lowest observed for CaMg$_2$Bi$_2$ ($\rho(300$\,K)/$\rho(10$\,K) = 1.2).  These results suggest that defect scattering significantly affects carrier transport in these materials, especially in the alkaline earth compounds.

While large single crystals of $A$Mg$_2$Bi$_2$ were grown from nominal melt compositions of $A$Mg$_4$Bi$_6$, the crystals were generally irregularly shaped and were not easily processed into the geometries required for thermoelectric transport measurements at high temperatures.  Therefore, polycrystalline samples were produced to evaluate the thermoelectric efficiency in these compounds.  However, the as-synthesized polycrystalline samples are found to contain different carrier densities than the single crystalline samples.  This discrepancy highlights the lack of knowledge regarding defects in this family of compounds.  It is particularly interesting that the polycrystalline samples have a lower carrier concentration, suggesting these polycrystalline materials contain fewer acceptor-like defects, or more electron donating impurities or compensating defects than the single crystals.

\section{Methods}
\subsection{Synthesis}

Polycrystalline samples of CaMg$_2$Bi$_2$, EuMg$_2$Bi$_2$, and YbMg$_2$Bi$_2$ were prepared via melting, grinding, and annealing.  In a He glove box, the elements were placed in a tantalum tube, which was then welded shut under an Ar atmosphere; the starting materials were 99.999\% Bi, 99.95\% Mg, 99.9\% dendritic Ca from Alfa, and high-purity Yb and Eu from Ames Laboratory. The tantalum tube was subsequently sealed within an evacuated quartz ampoule.  Two different melt procedures were employed: (1) Melting at 1000$^{\circ}$C for 84\,h.  (2) Melting at 1250$^{\circ}$C for 30-60\,min.  Samples produced via (1) utilized stoichiometric melts, while those synthesized using (2) employed 1\% excess of Ca, Yb, or Eu.  Following melting, samples were ground in a He glove box, cold pressed, and annealed under vacuum at 600$^{\circ}$C for at least 12\,h; this annealing step took place for all samples made via (1) and for the YbMg$_2$Bi$_2$ sample produced via (2).  No trend between the observed carrier concentration and the synthesis procedure was observed. In the following text, the synthesis route of a sample is distinguished by numbers corresponding to these procedures, and in the figures samples produced via (1) are represented by open markers and samples prepared via (2) are represented by closed markers.   For neutron diffraction measurements, a procedure similar to (2) was utilized followed by annealing at 600$^{\circ}$C for 10\,h.  For procedure (1), total masses near 15\,g were employed, while 5-7\,g was used for (2) and for neutron diffraction samples.

Single crystals were grown from a melt with nominal composition $A$Mg$_4$Bi$_2$; here, these samples are denoted by the synthesis procedure number 3.\cite{AMg2Bi2_Inorg} The excess Mg-Bi flux was removed via centrifugation at 650$^{\circ}$C, and the growth is believed to occur at temperatures lower than 750$^{\circ}$C during which time a cooling rate of $\approx$ 4.2$^{\circ}$C/min was employed.  Further details can be found in Ref. \citenum{AMg2Bi2_Inorg}. Large crystals were obtained with this method, and the crystals generally contain one flat face orthogonal to the \textit{c}-axis, and the side opposite this face is typically irregular.

To measure the transport properties, polycrystalline powders were hot pressed under dynamic vacuum with an applied pressure of $\sim$ 10,000\,psi.  The graphite furnace was heated to approximately 650\,$^{\circ}$C over an hour, at which point the temperature was held constant for 1-2\,h.  The samples were hand ground in a He glove box, and then loaded into a graphite die using grafoil to separate the sample from the graphite die, and remained under a He atmosphere until they could be quickly transferred to the hot-press.  The density of all samples was at least 97\% of the x-ray density.  After cutting, samples were kept under dynamic vacuum or in a helium glove box between measurements.

\subsection{Crystal structures and phase purity}

Crystal structures were characterized via x-ray and neutron powder diffraction.  Results obtained are consistent with those reported in Reference \citenum{AMg2Bi2_Inorg} for ground single crystals.  Powder x-ray diffraction (pXRD) data were collected at ambient conditions on a PANalytical X'Pert Pro MPD using a Cu K$_{\alpha,1}$ monochromator.  The main impurity phase is found to be elemental Bi, with volume fractions estimated to be less than approximately 5\% in all samples.

Neutron powder diffraction measurements were performed at a series of temperatures ranging from 10\,K to 973\,K, using the POWGEN time-of-flight diffractometer at the Spallation Neutron Source, Oak Ridge National Laboratory. Samples of CaMg$_2$Bi$_2$ and YbMg$_2$Bi$_2$ were ground into fine powders in a He glove box, and then loaded inside thin-walled vanadium containers.  EuMg$_2$Bi$_2$ was not studied due to absorption. Measurements were first performed above 300K, with unsealed containers in vacuum inside a radiative furnace. Trace oxygen amounts led to a small amount of MgO formation. Subsequent measurements below 300K were performed with the sealed sample container filled with a partial atmosphere of helium, and mounted on a closed-cycle refrigerator in vacuum. In the configuration used, the measurements covered $d$-spacings from $\sim$0.3 to 3.5\,\AA. A vanadium standard was measured to correct the efficiency of the raw detector counts.\cite{Huq2011} Rietveld refinements were carried out using the GSAS software package and the EXPGUI interface \cite{GSAS,EXPGUI} with $d$-spacings from $\sim$0.5 to 3.0\,\AA.

Sample homogeneity was investigated using a Hitachi TM-3000 tabletop microscope equipped with a Bruker Quantax 70 EDS system.  This revealed inclusions of MgO in all samples, which were not observed via powder x-ray diffraction except for a very small peak in the scan of sample Ca2 (Table \ref{tab:props}).  MgO inclusions with dimensions up to $\approx$ 50\,$\mu$m were observed, while the Bi inclusions observed were generally 1-5\,$\mu$m.  The extreme broadness of the Bi peak in x-ray diffraction suggests the Bi inclusions are small and/or strained, and perhaps many are smaller than the resolution of this SEM.  Oxidation of the surface was evident if the samples were not polished immediately prior to examination.

\subsection{Transport measurements}

Thermoelectric transport measurements below room temperature were performed in a Quantum Design Physical Property Measurement System (PPMS) using the Thermal Transport Option (TTO).  These measurements utilized gold coated copper leads attached to the samples via silver epoxy (H20E Epo-Tek), which requires a $\sim$30\,min cure at 100$^{\circ}$C.  Electrical resistivity and Hall effect measurements were performed using a standard four-point configuration in the PPMS; samples were thinned and electrical contacts were made with 0.025mm platinum wire and silver paste (DuPont 4929N).  Above room temperature, the Seebeck coefficient and electrical resistivity were measured in an ULVAC ZEM-3 M8; these samples were cut to rectangular bars between approximately 7.5 and 10\,mm long and 1.5-2\,mm on a side.  Thermal diffusivity $D_T$ measurements were conducted on $\sim$10\,mm diameter discs between 1.5 and 2\,mm thick, coated with graphite, in an Anter FL-5000 with a graphite furnace; data analysis followed ASTM 1461 for flash diffusivity.  The thermal conductivity $\kappa$ (above room temperature) was calculated via $\kappa = D_T \times C_P \times d$ where $d$ is the density. The specific heat capacity $C_p$ was approximated using the Dulong Petit limit $C_v$= 3R, calculated from the composition. Based on the equation $C_p=C_v+\alpha_T^2T/(d\beta_T$), where $\alpha_T$ is the coefficient of thermal expansion and $\beta_T$ is the isothermal compressibility, in the 300-600\,K temperature range, the thermal expansion term is small and will cause a slight underestimation of $C_p$.  Samples were lightly polished prior to measurements.

The large single crystals utilized to obtain thermoelectric transport data at low temperatures were not processed into standard geometries, and could not be measured at high temperatures.  As such, the mean values of $\rho$ for single crystals (see Table \ref{tab:props}) were utilized to obtain the transport dimensions and calculate the electrical resistivity and thermal conductivity.  The small distribution of properties observed in smaller, regularly shaped single crystals allows this method to be utilized without impacting the trends observed in $\rho$ and $\kappa$, though the absolute values may be off by up to $\sim$15\%.  

\subsection{First Principles Calculations}

Boltzmann transport calculations of the temperature and doping dependent Seebeck coefficients were performed for CaMg$_2$Bi$_2$, as well as CaMg$_2$Sb$_2$ and CaMg$_2$As$_2$ for comparison.  These calculations are based on the electronic structure obtained with the all-electron general potential linearized augmented planewave (LAPW) method,\cite{singh-book} using the WIEN2k code.\cite{wien} Very dense samplings of the Brillouin zone were used with well converged LAPW basis sets with local orbitals to relax linearization errors and include semicore states.\cite{singh-lo} The zone samplings were done with uniform 32$\times$32$\times$16 grids, while the cut-offs for the basis set $K_{\rm max}$, were obtained with the criterion $RK_{\rm max}$=9, where $R$ is the smallest LAPW sphere radius. The transport calculations were done within the constant scattering time approximation (CSTA), which assumes that the relaxation time is independent of energy, using the BoltzTraP code,\cite{boltztrap} as in our previous work on Pb chalcogenides.\cite{singh-pbte,parker-pbse}  This method gives predictions of $\alpha(T)$ as a function of doping with no adjustable parameters by assuming a rigid band model and shifting the chemical potential to simulate doping.

The approach for calculating the electronic structure is the same as that used in a prior study of halide scintillators with the TB-mBJ functional,\cite{singh-1} and the electronic structure of CaMg$_2$Bi$_2$ obtained with this method was already reported.\cite{AMg2Bi2_Inorg}  A standard generalized gradient approximation (GGA) functional is utilized to relax the internal coordinates in the unit cell by total energy minimization. This was done using the experimental lattice parameters from literature\cite{AMg2Bi21977} and the GGA of Perdew and co-workers (PBE).\cite{pbe} The resulting structural parameters were then used to calculate the electronic structures using the recently developed modified Becke Johnson functional of Tran and Blaha (TB-mBJ).\cite{mbj} While standard GGA functionals are designed to reproduce total energies and structures, but underestimate the band gaps of semiconductors,\cite{GGA} the TB-mBJ functional cannot be used for total energy calculations but instead yields very much improved band gaps in a wide variety of materials.\cite{mbj,singh-2} These calculations were performed relativistically, including spin orbit coupling for all elements. The LAPW sphere radii were 2.5 bohr for Ca and As, 2.6 bohr for Sb, and 2.8 bohr for Bi. The Mg LAPW sphere radius was chosen as 2.5 bohr, except for CaMg$_2$As$_2$ where a radius of 2.4 bohr was used.

\section{Results and Discussion}

The polycrystalline samples of $A$Mg$_2$Bi$_2$ produced via melting and hot pressing have carrier concentrations between 6$\times$10$^{18}$ and $\sim$1$\times$10$^{19}$holes/cm$^{3}$.  While $p$-type conduction is also observed in the single crystalline samples, the free carrier concentration is considerably lower in the polycrystalline samples.  These results are summarized in Table \ref{tab:props}, where standard deviations represent the distribution of properties observed for single crystals from measurements on multiple crystals as discussed in Reference \citenum{AMg2Bi2_Inorg}.  Here, the carrier concentration is characterized by the Hall carrier concentration, $n_H$ = 1/$R_He$ where $R_H$ is the Hall coefficient and $e$ the elemental charge.

\begin{table}[h]
\caption{Summary of room temperature properties in polycrystalline (P) and single crystalline (S) samples of $A$Mg$_2$Bi$_2$, as well as the effective mass calculated using a single parabolic band model; uncertainties for single crystal samples come from measurements on multiple crystals in Ref. \citenum{AMg2Bi2_Inorg}.  The sample IDs are utilized to identify different samples in the figures and the associated number is related to the synthesis procedure.}\label{tab:props}
\begin{tabular}{cccccccc}
\hline
 $A$ & P/S & ID & $n_H$  & $\mu_H$  & $\rho$  & $\alpha$ & $m^*$  \\
 - & - & -   &10$^{18}$cm$^{-3}$ & cm$^{2}$/V/s  & m$\Omega$ cm & $\mu$V/K & $m_e$ \\
\hline
Ca & P   & \ding{110},  Ca2  & 1.1  & 119   &  47.1 & 287  & 0.33  \\
Eu & P   & $\triangle$, Eu1  & 1.3  & 137   &  34.8 & 206  & 0.18 \\
Eu & P   & \ding{115},  Eu2  & 0.63 & 201   &  49.1 & 350  & 0.37 \\
Yb & P   & $\Circle$,   Yb1  & 2.9  & 131   &  16.3 & 240  & 0.42   \\
Yb & P   & $\CIRCLE$,   Yb2  & 8.3  & 123   &  6.1  & 190  & 0.54  \\
\hline
Ca & S & $\boxempty$, Ca3 & 20(3) &  105(15) & 3.05(25) &  115 & 0.46(4) \\
Eu & S & $\triangle$, Eu3 & 17(1) &  195(15) & 1.95(25) &   92 & 0.32(2)  \\
Yb & S & $\Circle$,   Yb3 & 46(6) &  175(25) & 0.85(5)  &   83 & 0.54(5)  \\ 
\hline
\end{tabular}
\end{table}

The source for $p$-type conduction is not known.  In single-crystalline $A$Mg$_2$Bi$_2$ and polycrystalline $A$Zn$_2$Sb$_2$ the rare-earth derived compounds have higher carrier mobilities, and the $A$=Yb compounds have higher carrier densities than the $A$=Eu,Ca compounds.  Given this correlation between $A$, the carrier density, and mobility, it seems reasonable that $A$ vacancies are the primary source for the charge carriers.  It is interesting that single crystals grown from an excess of Mg-Bi have higher carrier densities, and thus one suspects higher defect levels in the single crystals. The formation of $A$ vacancies may be promoted by growth in excess Mg-Bi, thus leading to larger defect concentrations.  Synthesis of polycrystalline samples from melts with 1\% excess $A$ did not result in systematic changes in $n_H$.

It is also possible that Ta incorporates into the polycrystalline samples, which would act as an electron donor and thus decrease the hole concentration.  While Ta was not detected via EDS, the concentration of Ta$^{5+}$ necessary to modify the carrier density is extremely small and is well below the resolution of EDS.  Finally, growth in a sealed vessel may result in higher Mg content, and thus Mg vacancies cannot be eliminated as a possible source for free carriers.    In the related $A$Zn$_2$Sb$_2$, the carrier concentrations are generally manipulated via isoelectronic substitutions, for example in solid solutions based on Zn-Cd substitutions or Ca-Yb substitutions.  Non-isoelectronic doping, such as La for Yb, typically fails to change the carrier concentration in these materials.  Thus, the defects in these materials warrant further investigation, and the unintentional manipulation of carrier density in these materials is perhaps one of the most important aspects of the current study.

\subsection{Low Temperature Transport Properties}

\begin{figure}
	\centering
\includegraphics[width=3in]{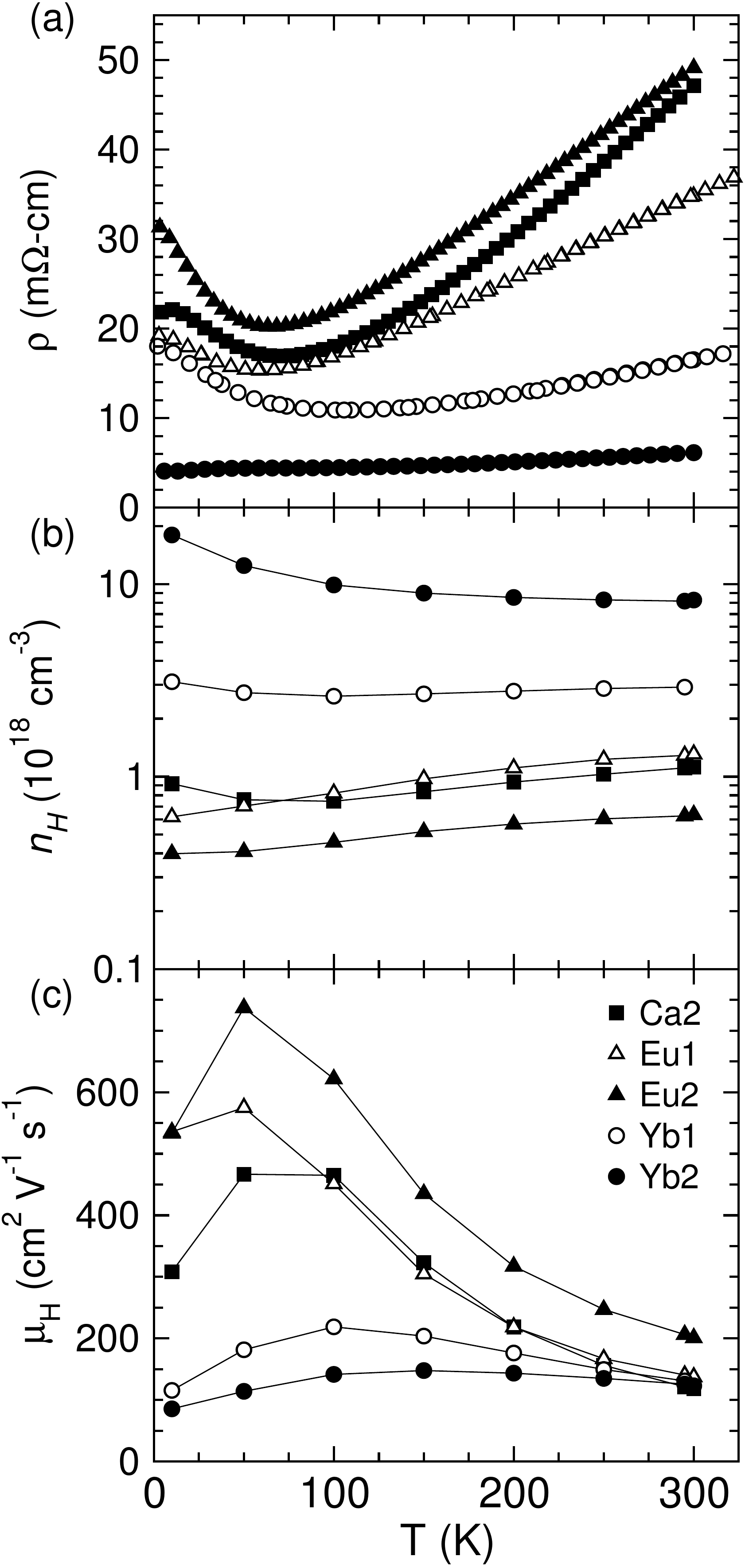}
\caption{(a) Electrical resistivity,  (b) Hall carrier density, and (c) Hall mobility of polycrystalline CaMg$_2$Bi$_2$, EuMg$_2$Bi$_2$, and YbMg$_2$Bi$_2$ samples with symbols linked to Table \ref{tab:props}.}
	\label{fig:Hall}
\end{figure}

The electrical resistivity $\rho$, Hall carrier density $n_H$, and Hall mobility ($\mu_H$=$R_H/\rho$) of polycrystalline samples are shown in Figure \ref{fig:Hall}.  The temperature dependence of the electrical resistivity and Hall mobility are similar in all samples, with a suppression of the mobility observed at low temperatures leading to an increase in the resistivity.  This behavior was not observed in single crystals of YbMg$_2$Bi$_2$ and EuMg$_2$Bi$_2$ (nor in sample Yb2), though sample Ca3 and other CaMg$_2$Bi$_2$ single crystals \cite{AMg2Bi2_Inorg} did possess qualitatively similar behavior.

\begin{figure}
	\centering
\includegraphics[width=3in]{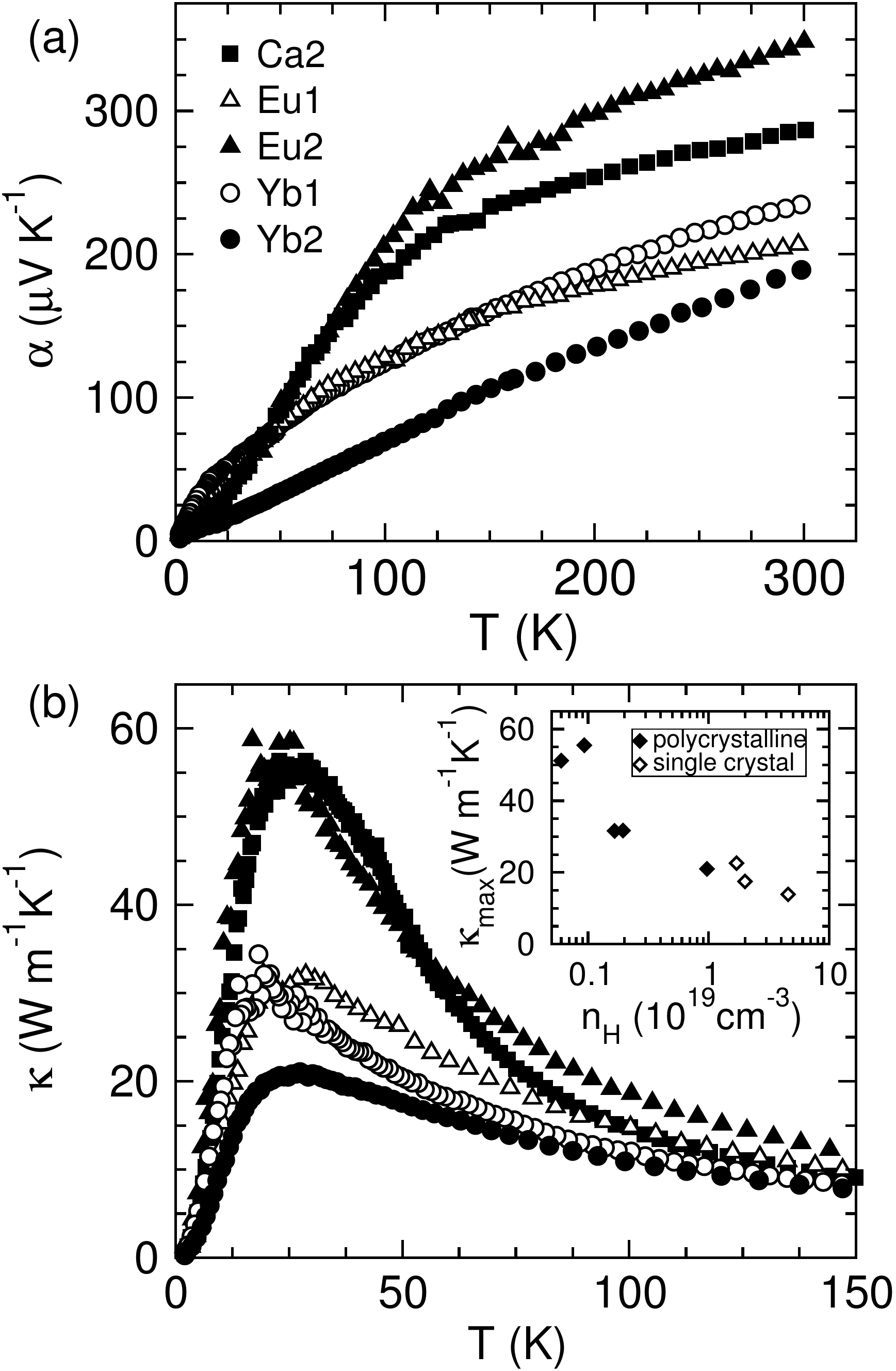}
\caption{The (a) Seebeck coefficient and (b) thermal conductivity in polycrystalline CaMg$_2$Bi$_2$, EuMg$_2$Bi$_2$, and YbMg$_2$Bi$_2$ below room temperature, with symbols linked to Table \ref{tab:props}.  The inset in (b) plots $\kappa$ at the low temperature peak versus the room temperature carrier concentration, and includes data for single crystalline samples.  The estimated lattice thermal conductivity is indistinguishable from the data in (b), as the largest estimate for the electronic contribution is less than 0.1\,W/m/K (for YbMg$_2$Bi$_2$ at 150\,K).}
	\label{fig:TTO}
\end{figure}

Given the uncertainty in defect identities and concentrations in these materials, one can only speculate on the source of decreased $\mu_H$ at low $T$.  The most plausible sources are the formation of a mobility edge due to disordered defects, the scattering of holes by point defects or ionized impurities, or activation through impurities (MgO) at the grain boundaries.  While activation through `dirty' grain boundaries offers a slightly simpler explanation, it does not account for the observation of similar behavior in single crystalline CaMg$_2$Bi$_2$.

\begin{figure}
	\centering
\includegraphics[width=3in]{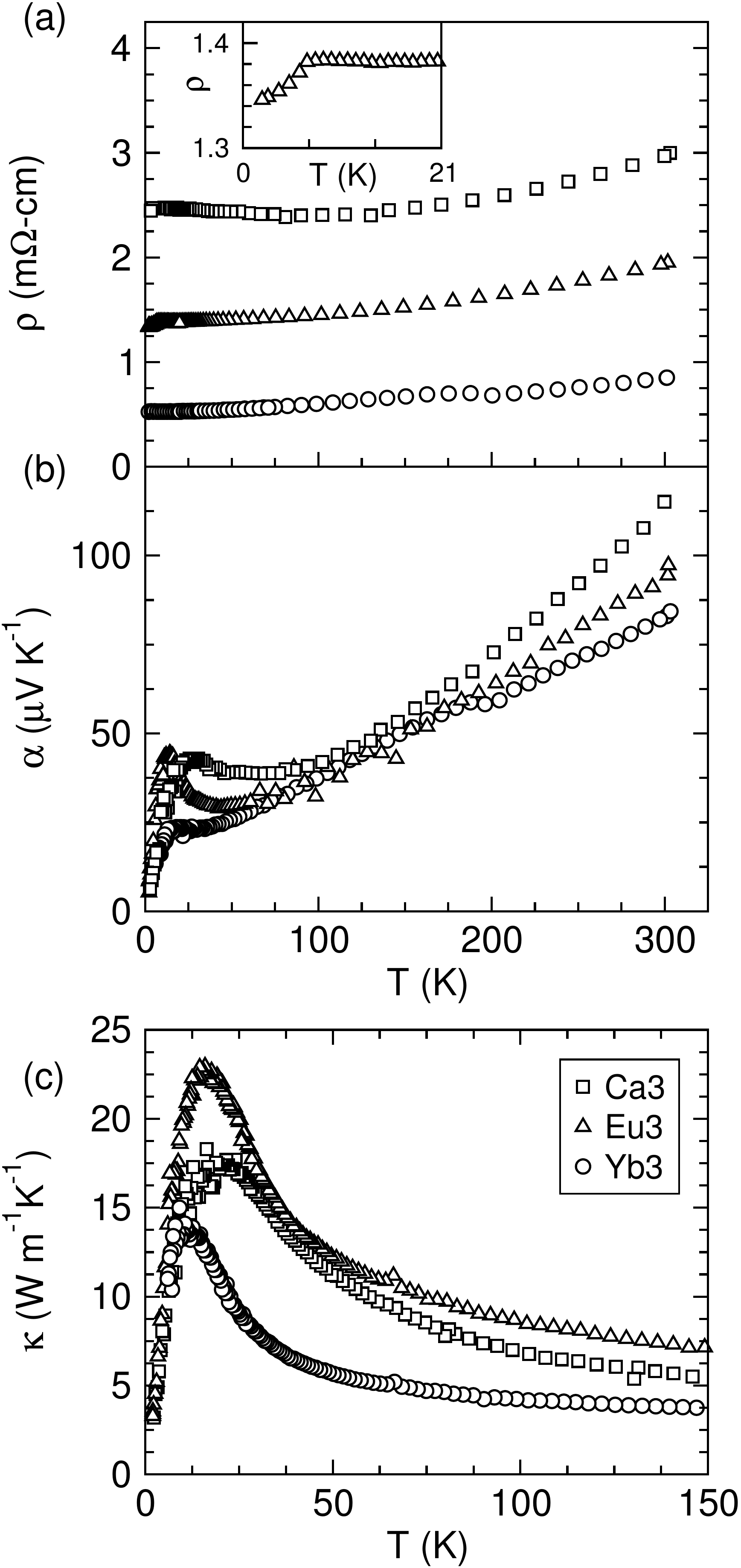}
\caption{(a) Electrical resistivity, (b) Seebeck coefficient, and (c) thermal conductivity in single crystals of CaMg$_2$Bi$_2$, EuMg$_2$Bi$_2$, and YbMg$_2$Bi$_2$ below room temperature. In (a) and (c) the electrical resistivity and thermal conductivity are scaled to match electrical resistivity data in smaller crystals (Table \ref{tab:props}) to account for the slightly irregular shape of the large single crystals.  The estimated lattice thermal conductivity is within the size of the data markers in (c); the largest electronic contribution to $\kappa$ is estimated to be 0.55W/m/K for YbMg$_2$Bi$_2$ at 150\,K.}
	\label{fig:xtls}
\end{figure}

Transport data for single crystals clearly show that defect scattering strongly influences the mobility in these compounds (low resistivity ratios implied by Figure \ref{fig:xtls}a).\cite{AMg2Bi2_Inorg} The scattering of carriers by point defects is typically modeled with a temperature independent relaxation time, which is generally associated with the residual resistance in a material.  When point defects are considered ionized impurities, however, the scattering typically results in a mobility that increases with increasing temperature as roughly $T^{1.5}$.\cite{Fistul} In the single crystals, defect scattering appears to be stronger in CaMg$_2$Bi$_2$ than in YbMg$_2$Bi$_2$ and EuMg$_2$Bi$_2$, and this is consistent with the trends observed in $\mu_H$ for polycrystalline $A$Zn$_2$Sb$_2$ samples at room temperature.  In these materials, $\rho$ is increasing with $T$, which implies that at high $T$ phonon scattering limits the carrier relaxation time.  When acoustic phonon scattering limits the carrier relaxation time, the mobility decreases with increasing temperature as $T^{-1.5}$ for non-degenerate carriers and as $T^{-1}$ for degenerate carriers.  

The transport properties of CaMg$_2$Bi$_2$ and EuMg$_2$Bi$_2$ reveal the importance of defect scattering, as a larger mobility is observed in the low-carrier concentration polycrystalline samples than in the high-carrier concentration single-crystalline samples (at low $T$).  To our knowledge, these low temperature $\mu_H$ values are the largest mobilities reported in CaAl$_2$Si$_2$ type materials.  The mobility in polycrystalline YbMg$_2$Bi$_2$ is suppressed below that of the single crystalline material despite a lower defect level (Table \ref{tab:props}), which suggests that grain boundary scattering may also be important.  However, it is important to note the irregular temperature dependence of $n_H$ for the YbMg$_2$Bi$_2$ samples and thus this $\mu_H$ data may be an artifact associated with the simple, single band model.

The thermoelectric transport properties below room temperature are shown in Figure \ref{fig:TTO} for polycrystalline samples and Figure \ref{fig:xtls} for single crystalline samples.  The lower carrier density in the polycrystalline samples is clearly reflected in the increased Seebeck coefficient.  The Seebeck coefficients decrease with decreasing temperature and increasing $n_H$, as expected for simple, doped semiconductors.

The inset in Figure \ref{fig:xtls}a shows the electrical resistivity of EuMg$_2$Bi$_2$ across the antiferromagnetic transition near 7\,K.\cite{AMg2Bi2_Inorg}.  While approaching this transition from higher temperatures, the electrical resistivity increases slightly then decreases sharply below the transition.  This behavior is associated with increased magnetic fluctuations near the transition leading to increased carrier scattering. The Seebeck coefficient also rises sharply and reaches a maximum before falling rapidly towards zero.  A large peak was observed in the heat capacity near this temperature, and analysis revealed an entropy change consistent with the magnetic ordering of Eu$^{2+}$ where $J=\frac{7}{2}$.\cite{AMg2Bi2_Inorg}  Interestingly, a similar, yet broad increase in $\alpha$ was observed in the single-crystalline CaMg$_2$Bi$_2$, possibly associated with changes in the dominant scattering mechanism.

The Seebeck coefficient and Hall carrier density can be utilized to obtain a density of states effective mass $m^*$ when a dispersion relationship and a carrier scattering model are assumed.  Given the decay of the carrier mobility with increasing temperature above roughly 100\,K, it is clear the carrier relaxation time is strongly influenced by acoustic phonon scattering.  It is very difficult to account for the influence of defect scattering on the relaxation time. Therefore, effective mass values are calculated assuming a simple model, with transport in a single parabolic band (SPB) that is limited by acoustic phonon scattering.  This SPB model analyzes the Hall carrier density $n_H=1/R_He$, which is equivalent to $n=r_H/R_He$ where the chemical doping level $n$ calculation utilizes the Hall factor $r_H$ to account for scattering effects that cause a difference between $n$ and $n_H$.  Further details of this SPB model can be found in Reference \citenum{BGGMay}.  

Within the SPB model, the effective masses of holes are between 0.2 and 0.6\,$m_e$ at room temperature, where $m_e$ is the free electron mass (see Table \ref{tab:props}).  For comparison, the Seebeck and Hall coefficients in $A$Zn$_2$Sb$_2$ are well described by a SPB model with $m^*$=0.56\,$m_e$ for $A$=Ca,Sr,Eu,Yb.   Some variation between compositions exists, as well as between the single crystalline and polycrystalline samples.  In addition to possible changes in the band structure with composition, the influence of defect scattering clearly changes with doping level and sample composition and this can lead to changes in the calculated $m^*$.  The validity of a single-carrier model is also brought into question for the lowest carrier concentrations due to the appearance of carrier activation at only moderately high temperatures, as is the assumption of a SPB given that electronic structure calculations show\cite{AMg2Bi2_Inorg} the valence band edge is dominated by two light bands (though they possess similar extrema).  A detailed theoretical study, such as that presented in Reference \citenum{ZincAntimonidesBo2011} would likely be beneficial for the development of a greater understanding of transport in these materials.

Due to the change in either conduction type or scattering mechanism at low temperature, the Hall coefficient is not analyzed in detail and the Hall carrier density $n_H$=1/$R_H/e$ is plotted in Figure \ref{fig:Hall}c. At room temperature, the data suggest $r_H$ varies between roughly 1.17 and 1.15, which is consistent with the low carrier concentrations.  A maximum of roughly 1.18 is obtained at the non-degenerate limit and thus the `true' or `chemical' carrier concentrations are likely higher than the Hall carrier concentration by about 16\%, which does not influence the general trends discussed.

The thermal conductivity data shown for polycrystalline samples (Figure \ref{fig:TTO}b) and  single crystalline samples (Figure \ref{fig:xtls}c) reveals that defect scattering also influences the thermal transport in these materials.  This behavior is highlighted in the inset of Figure \ref{fig:TTO}b, which plots the value of the low-temperature peak in $\kappa$ versus carrier concentration.  The decrease in this $\kappa_{max}$ is likely due to the point-defect scattering that is associated with hole formation.  The scattering of phonons by free carriers is unlikely in these materials due to low carrier concentrations and small band masses.\cite{ZimanEP1}  At higher temperatures, phonon-phonon scattering dominates the thermal transport and $\kappa$ decreases with increasing temperature.  The influence of defect scattering is observed through a comparison of Figure \ref{fig:TTO}b for polycrystalline samples and Figure \ref{fig:xtls}c for single crystalline samples, which reveals lower thermal conductivity in the single crystalline specimens (in the \textit{ab}--plane).  The single-crystalline data represents the high-carrier concentration markers in the inset of Figure \ref{fig:TTO}b.

\subsection{High Temperature Transport Properties}

\begin{figure}
	\centering
\includegraphics[width=3in]{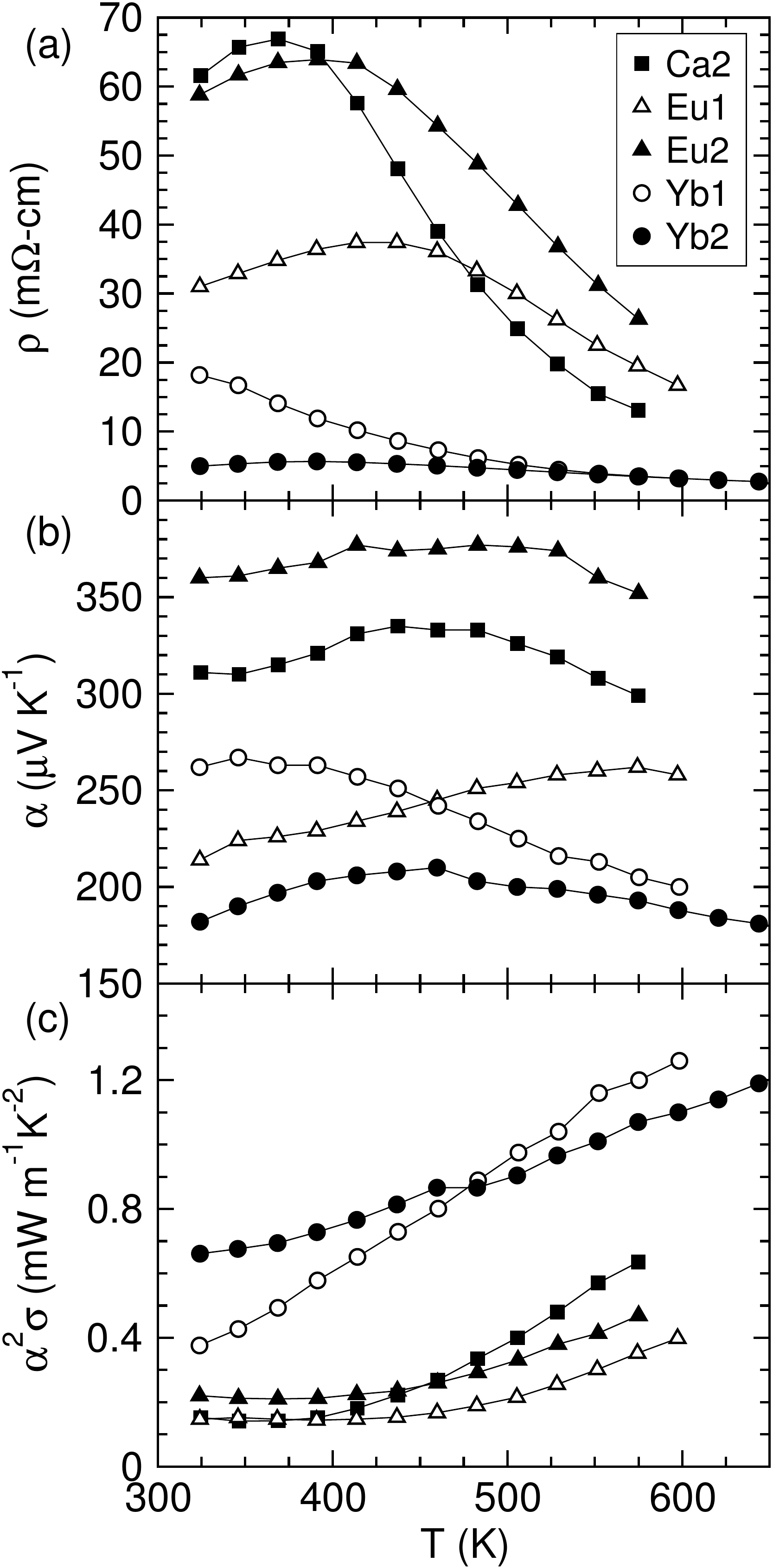}
\caption{High temperature (a) electrical resistivity,  (b) Seebeck coefficient, and (c) thermoelectric power factor of polycrystalline CaMg$_2$Bi$_2$, EuMg$_2$Bi$_2$, and YbMg$_2$Bi$_2$ samples with symbols linked to Table \ref{tab:props}.}
	\label{fig:ZEM}
\end{figure}

The high temperature electrical transport properties of polycrystalline samples are shown in Figure \ref{fig:ZEM}.  The electrical resistivity first rises with increasing $T$, due to phonon scattering, then reaches a maximum and decreases.   The observed maximum is associated with the thermal activation of charge carriers, and a corresponding maximum is observed in the Seebeck coefficients as well.  

Assuming charge carriers are activated across the energy gap, and not from impurity states, the temperature dependence of the resistivity above the maximum is expected to be described by $\rho = \rho_0$Exp[$(E_G)/(2kT)$], where $E_G$ is the thermal energy gap.  While limited data exist at these high temperatures, reasonable fits are obtained yielding 0.44\,eV for Ca2, 0.2\,eV for Yb2, and 0.3\,eV for Eu1, Eu2, and Yb1.  A band gap can also be estimated from the maximum in the Seebeck coefficient ($\alpha_{max}$ at temperature $T_{max}$) via $E_G = 2\alpha_{max}T_{max}e$.\cite{EgEstimate}  For these materials, this relationship yields energy gaps of roughly 0.2\,eV for the Ca-- and Eu--based materials, and $E_G \sim$0.3\,eV for YbMg$_2$Bi$_2$.  These estimates of $E_G$ are somewhat lower than those predicted by first principles calculations, which yielded a direct gap of 0.7\,eV for CaMg$_2$Bi$_2$.\cite{AMg2Bi2_Inorg}  

As expected from the low temperature data, the polycrystalline samples have large Seebeck coefficients above room temperature.  The Seebeck coefficients appear relatively independent of temperature while the electrical resistivity decreases with increasing temperature above $\sim$450\,K for all samples.  This is in part due to the scale necessary to view all of the data, as the Seebeck coefficients generally possess the expected temperature dependence; $\alpha$ first increases with $T$ then decreases at high $T$ consistent with $\rho(T)$. However, the temperature dependence of $\alpha$ and $\rho$ implies strong asymmetry in electron / hole conduction.

The relative rates of decrease in $\alpha$ and $\rho$ at high temperatures suggest that hole mobility is much larger than electron mobility.  This would cause the electronic properties to be dominated by hole conduction, and the decreases in $\alpha$ and $\rho$ would be primarily associated with the increase in hole concentration due to the activation of electron-hole pairs.  The thermoelectric power factor ($\alpha^2/\rho$ = $\alpha^2\sigma$ shown in Figure \ref{fig:ZEM}c) continues to increase with increasing temperature despite the appearance of intrinsic conduction, supporting the idea that a large asymmetry in electron-hole properties exists.

In a moderately doped semiconductor with symmetric band properties, the power factor decreases at high temperatures due to the compensation of the Seebeck coefficient from minority carriers; the decrease in $\rho$ is not enough to overcome the sharp decrease in $\alpha^2$.  However, deviation from this behavior occurs when the band properties are highly asymmetric.  In the current scenario, electronic structure calculations show that the density of states is much larger in the conduction band than in the valence band,\cite{AMg2Bi2_Inorg} which is associated with a larger effective mass for electrons.  The mobility is classically defined as being proportional to the ratio of the relaxation time to the effective mass, and the relaxation time itself is generally a function of the effective mass.  When acoustic phonon scattering limits the carrier relaxation times, the mobility is proportional to $m*^{-5/2}$ and the conductivity is proportional to $m*^{-1}$ (Ref. \citenum{Fistul}), thus an increase in the effective mass generally decreases the conductivity and relative contribution of that carrier to the transport properties.  Another possibility is that phase impurities or impurity states in the polycrystalline samples consume the thermally activated electrons.  In either case, the current data support the claim that hole transport dominates the electronic transport properties, even in the high-temperature intrinsic region.  This type of behavior is favorable for thermoelectric application.

\begin{figure}
	\centering
\includegraphics[width=3in]{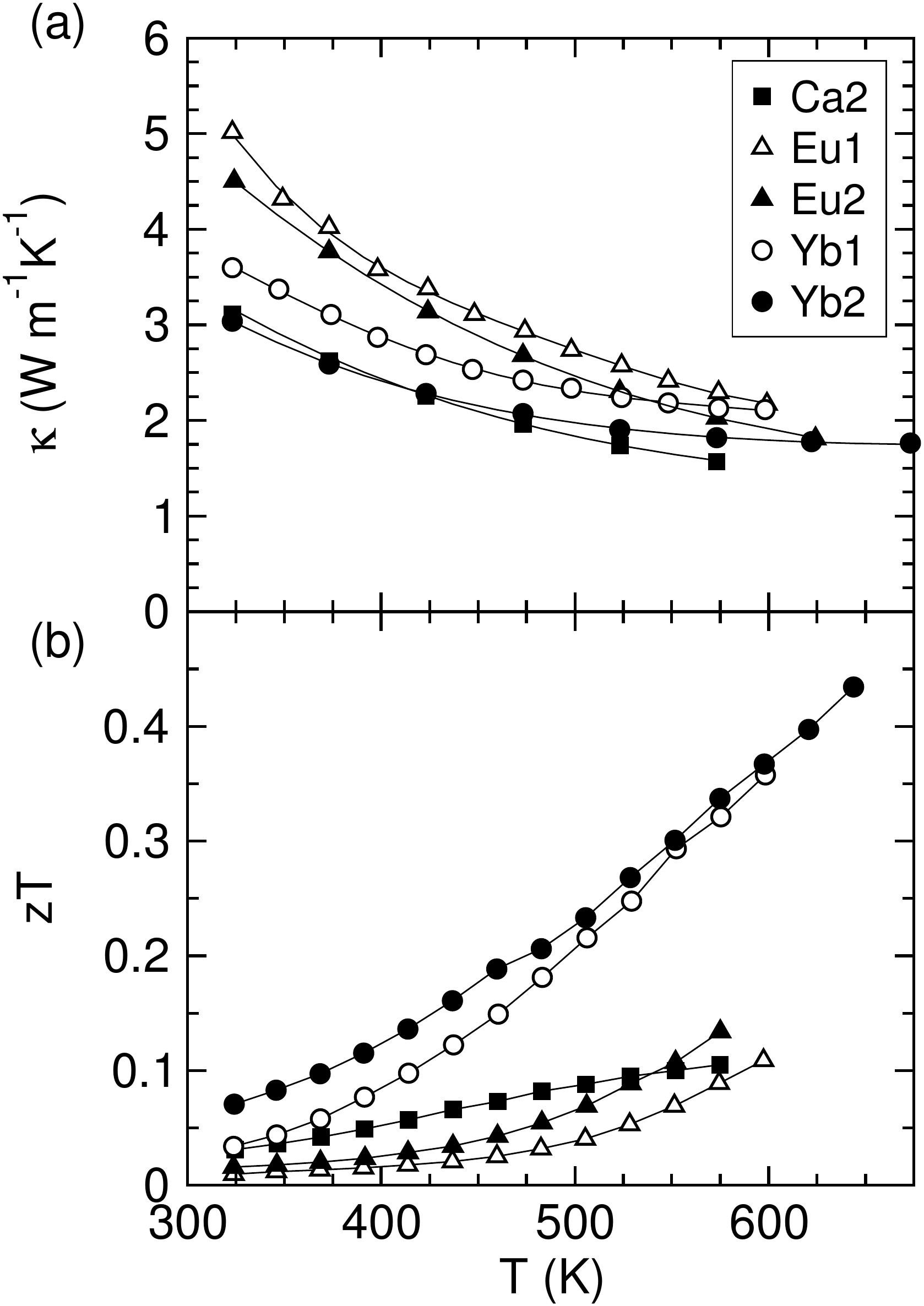}
\caption{(a) Thermal conductivity assuming a heat capacity of 3$RN$, (b) thermoelectric figure of merit $zT$ for polycrystalline CaMg$_2$Bi$_2$, EuMg$_2$Bi$_2$, and YbMg$_2$Bi$_2$.}
	\label{fig:zT}
\end{figure}

The high temperature thermal conductivity data are shown in Figure \ref{fig:zT}a.  A decrease in $\kappa$ with $T$ is observed, which is expected for these crystalline materials.  However, defect scattering does influence the temperature dependence and the decay is not as strong as the $T^{-1}$ expected for phonon-phonon scattering. Additional factors influence this temperature dependence, such as being near the minimum thermal conductivity.  Interestingly, the thermal conductivity is largest in EuMg$_2$Bi$_2$, and variations in the defect concentration do not trend with $\kappa$ at room temperature.  Electronic contributions are very low due to the large electrical resistivities, particularly at room temperature, and thus the differences in $\kappa$ between samples are either due to inherent properties or variations in sample quality.

The thermoelectric figure of merit $zT$ rises with increasing temperature for all samples for all temperatures probed, as shown in Figure \ref{fig:zT}b.  The largest $zT$ is obtained for YbMg$_2$Bi$_2$, with both Yb1 and Yb2 possessing $zT\sim$0.35 at 600\,K, and $zT>0.4$ is obtained by 640\,K.  This value is similar to that reported for isostructural Mg$_3$Sb$_2$, which possesses $zT\sim0.55$ at 675\,K.\cite{Mg3Bi2ICT03}  The compound Mg$_3$Bi$_2$ is reported to be metallic,\cite{Mg3SbBi,Mg3Bi2ICT03} however, and thus can be expected to possesses low thermoelectric efficiency.

The sample with higher carrier concentration, Yb2, has a larger power factor at lower temperatures, while the lower doped Yb1 sample has a larger power factor at high temperatures due to greater temperature dependence.  This leads to slightly larger $zT$ in sample Yb2 at lower temperatures, which suggests that higher doping levels will result in larger $zT$ and in particular larger average $zT$.  Even the oversimplified SPB model suggests that sample Yb2 is essentially optimally doped at room temperature, and the optimal doping level is expected to increase roughly as ($m^*T$)$^{3/2}$.\cite{Ioffe}  The optimization of $zT$ clearly needs to be done for performance at higher $T$, as $zT$ is small at 300\,K.  Therefore, the doping levels in these samples are considered low for moderate to high-temperature thermoelectric application, and the carrier densities found in single-crystalline samples are more likely to provide large thermoelectric performance at high temperatures.  However, the apparent asymmetry in electron/hole conduction leads to moderate thermoelectric performance in these samples, and little difference is observed in samples with significantly different carrier densities.  It would be interesting to examine samples of the analogous $A$Zn$_2$Sb$_2$ materials with low doping levels to see if these trends remain.

\subsection{First Principles Calculations}

First principles calculations were employed to explore the influence of pnictogen (Pn) on the electronic structure and the corresponding transport properties of CaMg$_2$Pn$_2$.  This is motivated, in part, by large thermoelectric performance in the $A$Zn$_2$Sb$_2$ and $A$Cd$_2$Sb$_2$ compounds.  CaMg$_2$Bi$_2$, CaMg$_2$Sb$_2$, and CaMg$_2$As$_2$ are calculated to be semiconductors, with the calculated band gaps decreasing with increasing atomic mass of the pnictide.  The calculated band gaps are given in Table \ref{tab:struct} along with the structural parameters that were used in the calculation.   The gaps are of charge transfer character but there is some covalency between Mg and pnictogen states.  The valence bands have mainly pnictogen $p$ character, but with an admixture of Mg, while the conduction bands have Mg character, mixed with Pn.  These features can be observed in the density of states for CaMg$_2$Bi$_2$, which have been published in Reference \citenum{AMg2Bi2_Inorg}.  These results show that the density of states in the conduction band is much larger than that in the valence band, as noted above, and help to explain the increase in $\alpha^2\sigma$ with increasing $T$ at high $T$.

\begin{figure}
\includegraphics[width=\columnwidth]{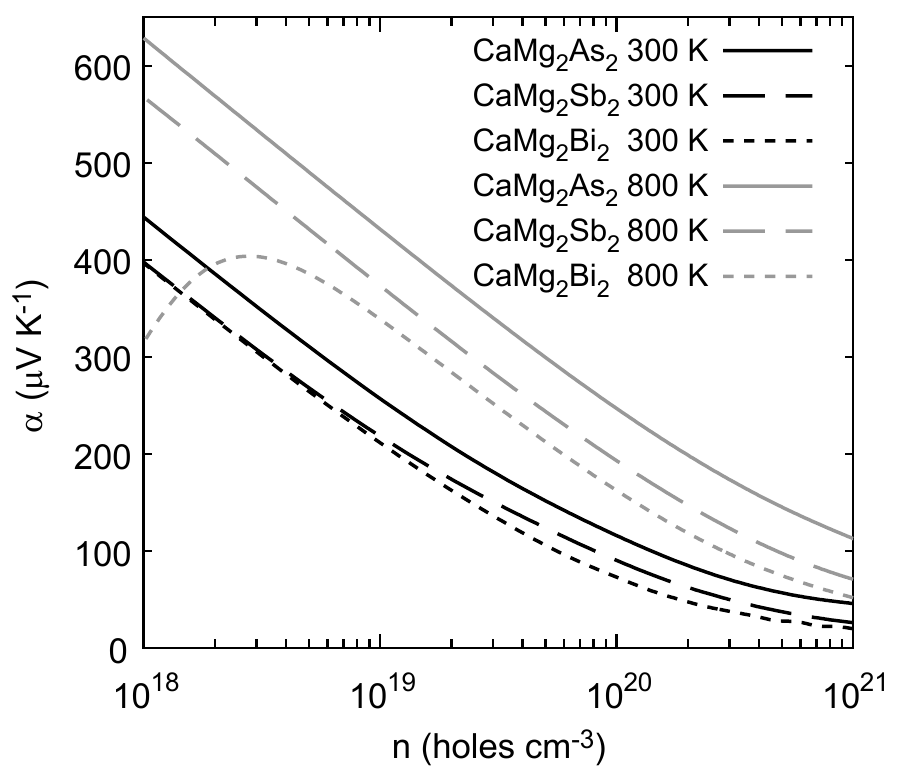}
\caption{Calculated Seebeck coefficient $\alpha$ for CaMg$_2$$Pn_2$, $Pn$=As,Sb,Bi, for $T$=300 K and 800 K. The quantity shown is the directional average of the Seebeck coefficient as described in the text.}
\label{fig:calc}
\end{figure}

The calculated Seebeck coefficients are shown as a function of $p$-type doping level in Fig. \ref{fig:calc} at 300\,K and 800\,K. The quantity shown is the linear average of the $c$-axis and the two equal in-plane $a$-axis values. This is an approximation to the ceramic average that is made because the anisotropy of the electrical and thermal conductivity are unknown.

As observed in Figure \ref{fig:calc}, the As and Sb compounds are predicted to have higher values of $\alpha$ for a given doping level, even when the doping level is given in units of cm$^{-3}$. Since the unit cell volume decreases as one goes from Bi to Sb to As, this trend in $\alpha(T)$ is even stronger if one considers the number of free carriers per unit cell, which is perhaps the more chemical approach.  The increase in $\alpha$ for a given $n$ is related to an increase in the effective mass.  This increase in band mass will reduce the carrier mobility, though the net effect on the power factor cannot be deduced.

The reduced band gap in CaMg$_2$Bi$_2$ relative to CaMg$_2$Sb$_2$ and CaMg$_2$As$_2$ results in a compensation of the Seebeck coefficient at high temperatures and low doping levels.  This is observed as the peak in $\alpha$ near $\sim$3$\times$10$^{18}$ holes per cm$^{3}$ for $T$=800\,K.  The influence of bipolar conduction is not predicted until higher temperatures for  CaMg$_2$Sb$_2$ and CaMg$_2$As$_2$. Clearly, the larger band gaps in the As and Sb-based compounds would facilitate operation at higher temperatures. 

Consistent with the above discussion, the magnitudes of the Seebeck coefficients suggest that the optimum doping level near room temperature is likely to be approximately 1$\times$10$^{19}$cm$^{-3}$, while operation at higher temperatures will require higher doping levels.  In the analogous antimonides, high $zT$ is obtained for Seebeck coefficients between approximately 180 and 250\,$\mu$V/K.\cite{Cd122Grin09,Cd122Grin10b,EuZn2Sb2,EuCdZnSb2}  In CaMg$_2$Bi$_2$, the calculation results qualitatively suggest optimal doping levels near 5$\times$10$^{19}$cm$^{-3}$ for operation at 800\,K, while the optimum doping is likely to be 1$\times$10$^{20}$cm$^{-3}$ or greater in the As and Sb compounds due to the increased band mass.

The changes in electronic structures imply that the As and Sb-based compounds may be interesting thermoelectric materials.  Also, these trends shed light on the way in which solid solutions may be used to tune the band properties.  A detailed investigation into the effects of band mass on the power factor is warranted, as is an investigation of the influence the pnictogen has on the thermal conductivity.

\begin{table}
\caption{Structural parameters used in the present calculations.  The lattice parameters $a$ and $c$ are experimental data from the literature, while the internal coordinates are determined by
energy minimization. The band gaps, $E_g$ are obtained with the TB-mBJ functional.}
\label{tab:struct}
\begin{tabular}{lccccc}
\hline \hline
compound~~~ & ~$a$(\AA)~ & ~$c$(\AA)~ & $z_{\rm Mg}$ & $z_{Pn}$ & $E_g$(eV) \\
\hline
CaMg$_2$As$_2$ & 4.34 & 7.13 & 0.6333 & 0.2481 & 1.77 \\
CaMg$_2$Sb$_2$ & 4.66 & 7.58 & 0.6303 &  0.2434 & 1.10 \\
CaMg$_2$Bi$_2$ & 4.73 & 7.68 & 0.6271 & 0.2382 & 0.70 \\
\hline
\end{tabular}
\end{table}

\subsection{Thermal Expansion}

Neutron powder diffraction was undertaken to examine the stability and expansion of the lattice. The source for free carriers remains unknown, with cation vacancies being one of the most plausible explanations. In YbMg$_2$Bi$_2$,  1 cation vacancy per 1500 unit cells would produce 1$\times$10$^{19}$holes/cm$^{3}$, and thus it is not possible to detect these defects via neutron powder diffraction.  Similarly, previous single crystalline x-ray diffraction could not isolate defects.\cite{AMg2Bi2_Inorg}

\begin{figure}
	\centering
\includegraphics[width=3in]{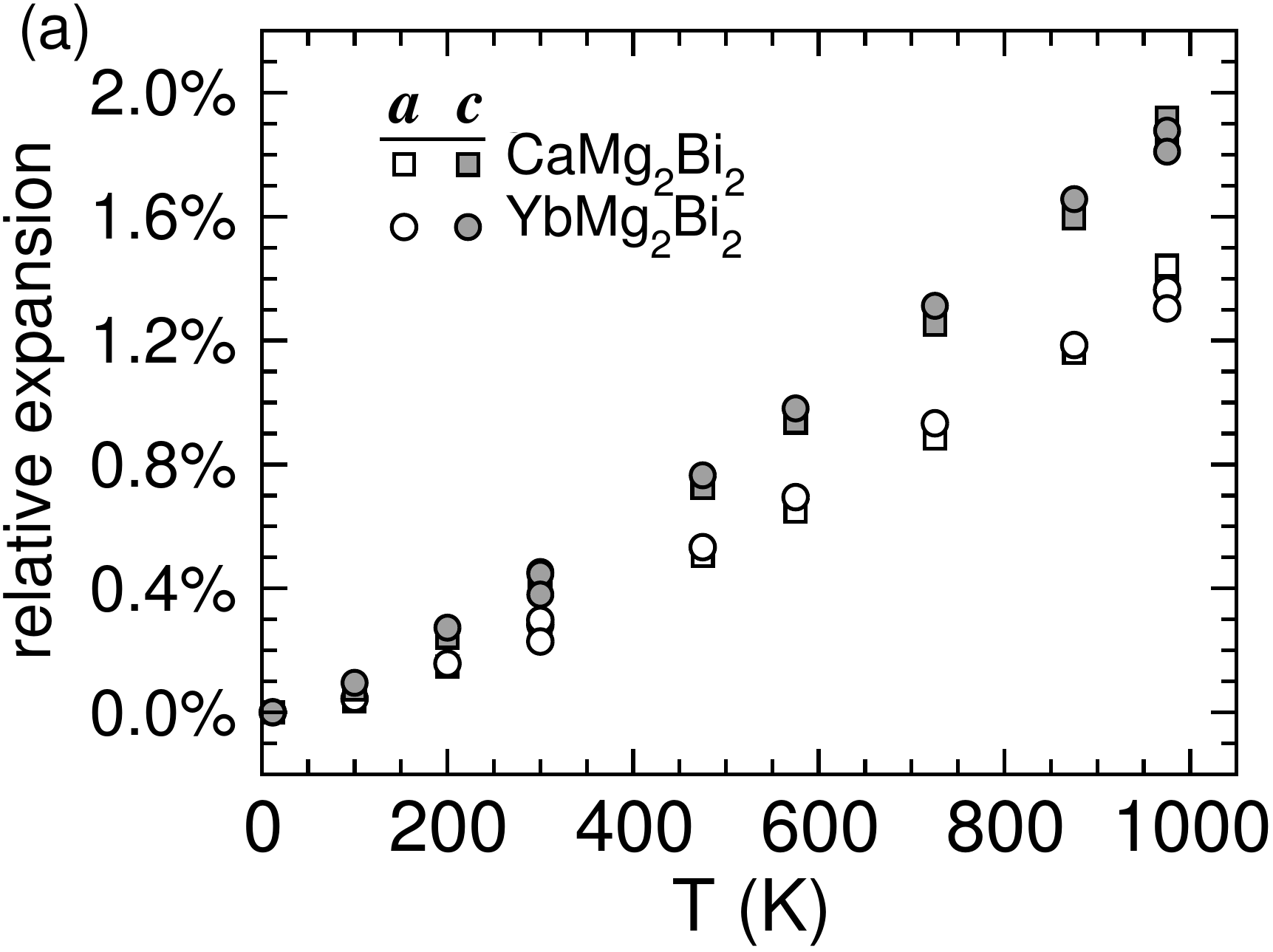}
\includegraphics[width=3in]{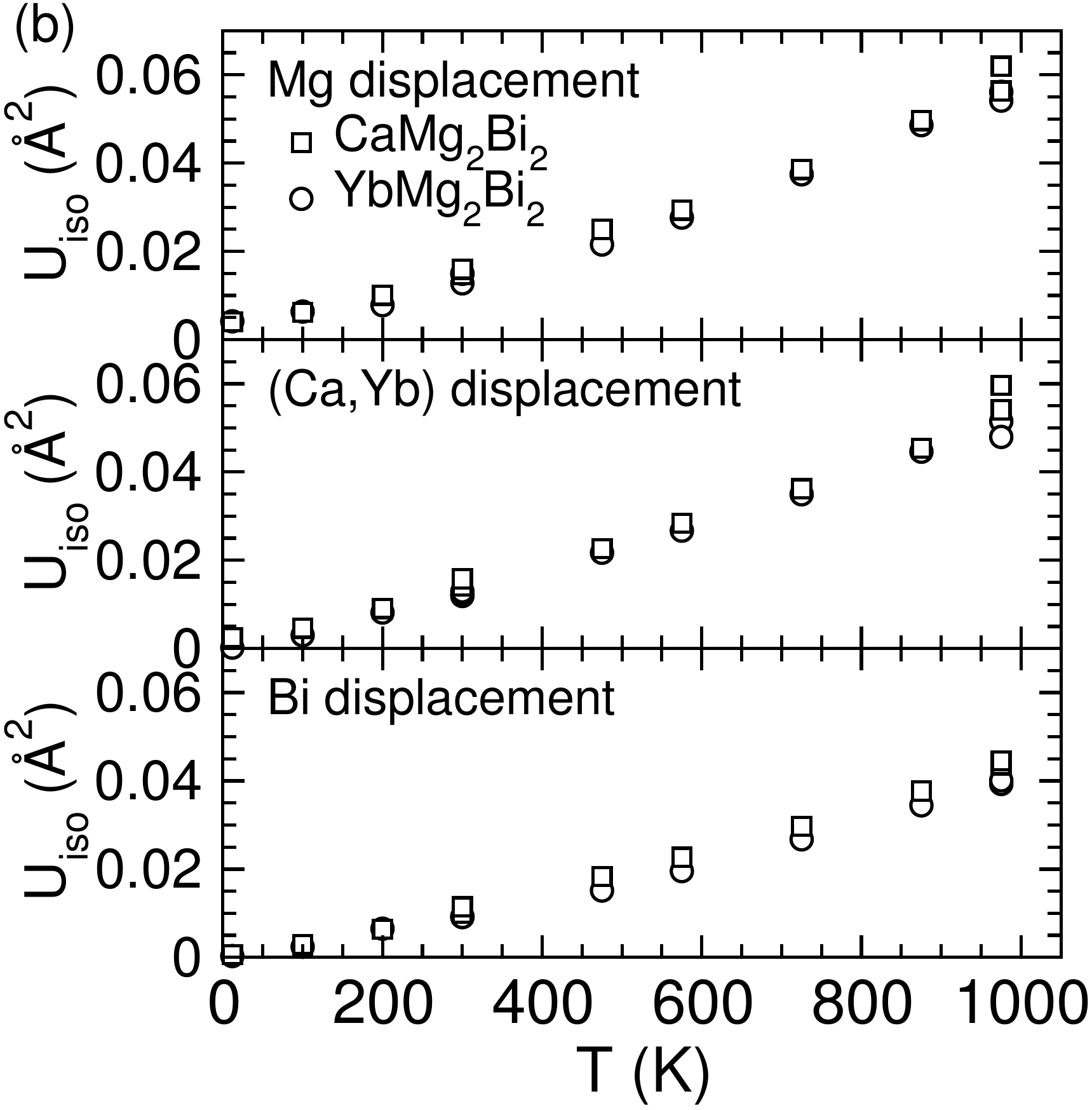}
\caption{Refinement results for neutron powder diffraction data showing the (a) relative change in lattice parameters and (b) the associated change in isotropic displacement parameters.  At 300\,K, the lattice parameters were \textit{a}=4.7326(1)\AA\, and \textit{c}=7.6780(2)\AA\, for CaMg$_2$Bi$_2$, and \textit{a}=4.7321(1)\AA\, and \textit{c}=7.6649(1)\AA\, for YbMg$_2$Bi$_2$.}
\label{fig:powgen}
\end{figure}

Figure \ref{fig:powgen}a presents the temperature dependence of the lattice parameters obtained from neutron powder diffraction.  The expansion of \textit{a} and \textit{c} is similar in both CaMg$_2$Bi$_2$ and YbMg$_2$Bi$_2$, as are the values and relative changes of the displacement parameters (Fig. \ref{fig:powgen}b).  The refinements possessed R$_{wp}$ of 0.0286 and 0.0282, and $\chi^2$ of 5.86 and 5.18 for CaMg$_2$Bi$_2$ and YbMg$_2$Bi$_2$, respectively, at 300\,K.  The structural parameters obtained are similar to those reported in Ref. \citenum{AMg2Bi2_Inorg} for ground single-crystals.

Thermal expansion is slightly larger along \textit{c} for both compounds, which is consistent with the first-level approximation of an anionic (Mg$_2$Bi$_2$)$^{2-}$ layer being supported by charge transfer from the cationic layer of $A^{2+}$.\cite{CaAl2Si2Bonds}  Figure \ref{fig:powgen} also reveals these compounds to be stable to nearly 1000\,K, which is consistent with previous differential thermal analysis to 1073\,K.  Minor oxidation occurred during the measurements as observed via the formation of MgO, the concentration of which increased as the measurement temperature increased reaching 2-4\% at 973\,K.  MgO was also observed in the polycrystalline samples after hot-pressing, and the surfaces of hot-pressed samples oxidized during the high-temperature measurements.  Thus, while these materials appear thermodynamically stable to high temperatures, they are readily oxidized and care must be taken to avoid catastrophic oxidation, which occurred during one measurement to 500$^{\circ}$C in the Anter system.  YbMg$_2$Bi$_2$ survived this measurement in the Anter system and appears less prone to oxidation than CaMg$_2$Bi$_2$ and EuMg$_2$Bi$_2$.

\section{Conclusion}

The low carrier density observed in polycrystalline CaMg$_2$Bi$_2$, EuMg$_2$Bi$_2$, and YbMg$_2$Bi$_2$ results in only moderate thermoelectric efficiency by 600\,K.  Two samples of YbMg$_2$Bi$_2$ possessed the most promising performance, with $zT\sim$0.4 obtained near 625\,K; these samples possessed the highest carrier densities of the polycrystalline materials.  The thermoelectric efficiency is likely to be larger at higher carrier concentrations, with the doping levels in the current polycrystalline samples being an order of magnitude or more lower than the expected optimum doping levels.  First principles calculations reveal a decreasing band gap and effective mass as the pnictogen is changed from As to Sb to Bi, behavior which could possibly be utilized to tune band properties and enhance thermoelectric performance.  The temperature dependences of the electrical resistivities and Seebeck coefficients suggest that the minority carriers activated at high temperatures have little impact on the electrical properties, likely due to large minority carrier effective mass.  Understanding the role of minority carriers and the defects responsible for the extrinsic carriers is critical for improving the thermoelectric efficiency of these and related compounds.

\section{Acknowledgements}

The high-temperature transport measurements were performed on instruments supported by the High Temperature Materials Laboratory User program via DOE EERE Office of Vehicle Technologies.  This work was also supported by the U. S. Department of Energy, Office of Basic Energy Sciences, Materials Sciences and Engineering Division (A.F.M., M.A.M., D.J.S.). The research at Oak Ridge National Laboratory's Spallation Neutron Source was sponsored by the Scientific User Facilities Division, Office of Basic Energy Sciences, U.S. Department of Energy.  O.D. and J.M. were supported by the U. S. Department of Energy, Office of Basic Energy Sciences, through the S3TEC Energy Frontier Research Center, Department of Energy DESC0001299. W.C. was supported by the Oak Ridge Associated Universities through ORISE's postdoctoral research program.  Use of the Spallation Neutron Source is also supported by the Division of Scientific User Facilities, Office of Basic Energy Sciences, U.S. Department of Energy, under contract DE-AC05-00OR22725 with UT-Battelle, LLC.

\end{document}